\documentclass[12pt]{article}
\usepackage{latexsym,amssymb}
\usepackage[american]{babel}
\makeatletter
\renewcommand{\@evenhead}{\raisebox{0pt}[\headheight][0pt]{\vbox{\hbox
to \textwidth{\thepage\hfil\strut\textsc{\leftmark}}\hrule}}}
\renewcommand{\@oddhead}{\raisebox{0pt}[\headheight][0pt]{\vbox{\hbox
to \textwidth{\textsc{\rightmark}\hfil\strut\thepage}\hrule}}}
\makeatother

\def\II{{\mathbb I}}
\def\RR{{\mathbb R}}
\def\CC{{\mathbb C}}

\def\ZZ{{\mathbb Z}}

\def\Id{{\rm \,Id\,}}

\def\dist{{\rm \,dist\,}}

\def\tr{{\rm \,tr\,}}
\def\Tr{{\rm \,Tr\,}}

\def\det{{\rm \,det\,}}

\def\End{{\rm \,End\,}}
\def\erf{{\rm \,erf\,}}
\def\erfc{{\rm \,erfc\,}}

\def\log{{\rm \,log\,}}

\def\vol{{\rm \,vol\,}}

\def\const{{\rm \,const\,}}
\def\be{\begin{equation}}
\def\ee{\end{equation}}
\def\bea{\begin{eqnarray}}
\def\eea{\end{eqnarray}}

\newtheorem{rem}{Remark}

\begin{document}

\begin{titlepage} 
\null

\hspace*{52truemm}{\hrulefill}\par\vskip-4truemm\par 
\hspace*{52truemm}{\hrulefill}\par\vskip5mm\par 
\hspace*{52truemm}{{\large\sc New Mexico Tech {\rm (October,
2001)}}}\vskip4mm\par 
\hspace*{52truemm}{\hrulefill}\par\vskip-4truemm\par 
\hspace*{52truemm}{\hrulefill} 
\par 
\bigskip 
\bigskip
\par
\vspace{3cm}
\centerline{\LARGE\bf Heat Kernel Asymptotics of} 
\medskip 
\centerline{\LARGE\bf Zaremba Boundary Value Problem} 
\bigskip 
\bigskip 
\centerline{\Large\bf Ivan G. Avramidi} 
\bigskip 
\centerline{\it Department of Mathematics} 
\centerline{\it New Mexico Institute of Mining and Technology} 
\centerline{\it Socorro, NM 87801, USA} 
\centerline{\it E-mail: iavramid@nmt.edu} 
\bigskip 
\medskip 
\vfill 
 
{\narrower
\par
The Zaremba boundary-value problem is a boundary value problem for Laplace-type
second-order partial differential operators acting on smooth sections of a
vector bundle over a smooth compact Riemannian manifold with smooth
boundary but with non-smooth (singular) boundary  conditions, which
include Dirichlet conditions on one part of the boundary and Neumann ones on
another part of the boundary. We study the heat kernel asymptotics of
Zaremba boundary value problem. The construction of the global parametrix of
the heat equation is described in detail and the leading parametrix is computed
explicitly. Some of the first non-trivial coefficients of the heat kernel
asymptotic expansion are computed explicitly.
\par} 
\vfill 
\end{titlepage} 

\section{Introduction}

The heat kernel of elliptic partial differential operators acting on 
sections of
vector bundles over compact manifolds proved to be of great importance in
mathematical physics. In particular, the main objects of interest in quantum
field theory and statistical physics, such as the effective action, the partition
function, Green functions, and correlation functions, are described by the
functional determinants and the resolvent of differential operators, which can
be  expressed in terms of the heat kernel. The most important operators
appearing in physics and geometry are the second order partial differential
operators of {\it Laplace type}; such operators are characterized by a scalar
leading symbol (even if acting on sections of vector bundles). Within the
\hbox{smooth} category this problem has been studied extensively during last
years (see, for example, \cite{gilkey95,berline92}; for reviews see
\cite{avramidi96,avramidi99,avramidi98}  and references therein). 
In the case of smooth compact manifolds without boundary the problem of
calculation of heat kernel asymptotics reduces to a purely computational
(algebraic) one for which various powerful algorithms have been developed
\cite{avramidi91b,vandeven98}; this problem is now well understood. In the
case of smooth compact manifolds with a smooth boundary and smooth boundary
conditions the complexity of the problem depends significantly on the type of
the boundary conditions. The classical smooth boundary problems (Dirichlet,
Neumann, or a mixed combination of those on vector bundles) are the most
extensively studied ones (see \cite{branson90a,branson99,kirsten98,avramidi93}
and the references therein). A more general scheme, so called oblique (or 
Grubb-Gilkey-Smith) boundary value problem \cite{grubb74,gilkey83b,gilkey83a},
which includes tangential (oblique) derivatives along the boundary,  has been
studied in
\cite{avramidi99a,avramidi99b,avramidi98b,dowker97,dowker98,elizalde98}.  In
this case the problem is not automatically elliptic; there is a certain strong
ellipticity condition on the leading symbol of the boundary operator. This
problem is much more difficult to handle, the main reason being that the heat
kernel asymptotics are no longer polynomial in the jets of the symbols of the
differential operator and the boundary operator. Another class of boundary
value problems are characterized by  essentially non-local boundary
conditions, for example, the spectral or Atiyah-Patodi-Singer boundary
conditions \cite{gilkey95,grubb96,booss-bavnbek93,seeley69a}.

All the above described boundary value problems were smooth. A more general
(and much more complicated) setting, so called {\it singular boundary
value problem}, arises when either the symbol of the differential operator or
the symbol of  the boundary operator (or the boundary itself) are not smooth.
In this paper we study a singular boundary value problem for a second order
partial differential operator of Laplace type when  the operator itself has
smooth coefficients but the {\it boundary operator is not smooth}. For the
case when the manifold as well as the boundary are smooth, but the boundary
operator jumps from Dirichlet to Neumann on the boundary, is known in the
literature as {\it Zaremba problem}.  Such problems often arise in  applied
mathematics and engineering and there are some exact results available for
special cases (two or three dimensions,  specific geometry, etc.)
\cite{sneddon66,fabrikant91}. Zaremba problem belongs to a much wider
class of singular boundary value problems, i.e. manifolds with singularities
(corners, edges, cones etc.). There is a large body of literature on this
subject where the problem is studied from an abstract function-analytical
point of view
\cite{fedosov64,cheeger79,cheeger83,callias83,bruening85,bruening91,simanca87,
mooers96,lesch97,karol98,gil01}.
However, the study of heat kernel asymptotics of Zaremba type problems is
quite new, and there are only some preliminary results in this area
\cite{avramidi00,dowker00a,dowker00b}. Moreover, compared to the smooth
category the needed machinery is still underdeveloped. We would like to stress
that we are interested not only in the  asymptotics of the trace of the heat
kernel, i.e. the integrated heat kernel  diagonal, but also in the local {\it
asymptotic expansion of the  off-diagonal heat kernel}.

This paper is organized as follows. In the sect. 2 a formal description of the
Zaremba problem is given. In the sect. 3 the general form of the heat kernel
asymptotic expansion is described. In sect. 4 the construction of the global
parametrix to the heat equation is described. In sect. 5 the first
non-trivial heat kernel coefficients are computed explicitly. In conclusion
we summarize (and discuss) the results and outline some future problems.

\section{General Setup}

\subsection{Laplace Type Operators}
%

Let $(M,g)$  be a smooth compact Riemannian manifold of dimension $m$ with a
boundary $\partial M$, equipped with a positive definite Riemannian metric
$g$. Let $V$ be a vector bundle over $M$, $V^*$ be its dual, and $\End(V)\cong
V\otimes V^*$ be the corresponding bundle of endomorphisms. Given any vector
bundle $V$, we denote by $C^\infty(V)$ its space of smooth sections.  We
assume that the vector bundle $V$ is equipped with a Hermitian metric.  This
naturally identifies the dual vector bundle $V^*$ with $V$, and defines a
natural $L^2$ inner product and the  $L^2$-trace using the invariant
Riemannian measure $d\vol_g$ on the manifold $M$.  The completion of
$C^\infty(V)$ in this norm defines the Hilbert space $L^2(V)$ of square
integrable sections.

We denote by $TM$ and $T^*M$ the tangent and cotangent bundles of $M$. Let a 
connection, $\nabla^V: C^\infty(V)\to C^\infty(T^*M\otimes V)$, on the vector
bundle $V$ be given,  which we assume to be compatible with the Hermitian
metric on the vector bundle $V$. The connection is given its unique natural
extension to bundles in the tensor algebra over $V$ and $V^*$.  In fact, using
the Levi-Civita connection $\nabla^{\rm LC}$ of the metric $g$ together with
$\nabla^V$, we naturally obtain connections on all bundles in the tensor
algebra over $V,\,V^*,\,TM$ and $T^*M$; the resulting connection will usually
be denoted just by $\nabla$.  It is usually clear which bundle's connection is
being referred to, from the nature of the section being acted upon. We also
adopt the Einstein convention and sum over repeated indices. With our
notation, Greek indices, $\mu,\nu,\dots$, label the local coordinates on
$M$ and range from 1 through $m$, lower case  Latin indices from the middle of
the alphabet, $i,j,k,l,\dots$, label the local coordinates on $\partial M$
(codimension one manifold) and range from $2$ through $m$, and lower case
Latin indices from the beginning of the alphabet, $a,b,c,d, \dots$, label the
local coordinates on a codimension two manifold $\Sigma_0\subset \partial M$
that will be described later and range over $3,\dots,m$. Further,  we will
denote by $\hat g$ the induced metric on the submanifolds (of the codimension
one or two) and by $\hat \nabla$ the Levi-Civita connection of the  induced
metric. We should stress from the beginning that we slightly abuse the
notation by using the same symbols for all submanifolds (of codimension one
and two). This should not cause any misunderstanding since it is always clear
from the context what is meant.

Let $\nabla^*$ be the formal adjoint of the covariant derivative defined
using the Riemannian metric and the Hermitian structure  on $V$ and let $Q\in
C^\infty(\End(V))$ be a smooth Hermitian section of the endomorphism bundle
$\End(V)$. The \textit{Laplace type operator} $F: C^\infty(V)\to C^\infty(V)$
is a partial differential operator of the form 
\be
F=\nabla^*\nabla+Q=-g^{\mu\nu}\nabla_\mu\nabla_\nu+Q\,.
\label{1ms}
\ee
Alternatively, the Laplace type operators are  second-order partial
differential operators with positive definite \textit{scalar leading symbol}
of the form  $\sigma_L(F;x,\xi)=\II\,|\xi|^2 = \II\,
g^{\mu\nu}(x)\xi_\mu\xi_\nu$.  Hereafter $\II$ denotes the identity
endomorphism of the vector bundle $V$. We will often omit it whenever it does
not cause any misunderstanding. Any  second order operator with a scalar
leading symbol can be put in the form (\ref{1ms}) by choosing the Riemannian
metric $g$, the connection $\nabla^V$ on the vector bundle $V$ and the
endomorphism $Q$.

\subsection{Boundary Conditions}

In the case of manifolds with boundary, one has to impose some boundary 
conditions in order to make a (formally self-adjoint) differential operator
self-adjoint (at least symmetric) and elliptic.
Let $N$ be the inward-pointing unit normal vector field to the boundary and
let $W=V\big|_{\partial M}$ be the restriction of the vector bundle $V$ to
the boundary $\partial M$. We define the {\it boundary data map}  $\psi:\
C^\infty(V)\to L^2(W\oplus W)$ by  
\be
\psi(\varphi)=\left(\matrix{\varphi|_{\partial M}\cr
\nabla_N\varphi|_{\partial M}\cr}\right)\,. \label{2}  
\ee
The boundary conditions then read 
%
%
\be 
B\psi(\varphi)=0\,,
\label{3} 
\ee
where $B: L^2(W\oplus W)\to L^2(W\oplus W)$ is the {\it boundary operator},
which will be specified later. If the operator $B$ is a tangential {\it
differential} operator (possibly of order zero), then the boundary
conditions are {\it local}. Otherwise, for example, when $B$ is a
pseudo-differential operator, the boundary conditions are {\it non-local}.
%

To define the boundary operator one needs a self-adjoint orthogonal projector
$\Pi$ that splits the space $L^2(W)$ in two orthogonal subspaces 
\be
L^2(W)=L_{||}^2(W)\oplus L_{\perp}^2(W)\,,
\ee
where
\be
L_{||}^2(W)=\Pi L^2(W)\,, \qquad {\rm and}\qquad 
L^2_{\perp}=(\Id-\Pi)L^2(W)\,,
\ee
and a {\it self-adjoint} operator $\Lambda:\ L^2(W)\to L^2(W)$,  such that
$\Lambda L^2_{||}(W)=\{0\}$, i.e.  $\Pi\Lambda=\Lambda\Pi=0$.
Hereafter $\Id$ denotes the identity operator. The boundary operator is then
defined by
\be 
B=\left(\matrix{\Pi&0\cr
\Lambda & \Id -\Pi\cr}\right)\,,
\label{5ms}
\ee
which is equivalent to the following boundary conditions
\bea
&&\Pi\left(\varphi\Big|_{\partial M}\right)=0\,,
\\
&&(\Id-\Pi)\left(\nabla_N\varphi\Big|_{\partial M}\right)
+\Lambda\left(\varphi\Big|_{\partial M}\right)=0\,,
\eea
It is easy to see that the boundary operator $B$ and the operator 
\be
K=\Id-B=\left(\matrix{\Id-\Pi&0\cr
-\Lambda & \Pi\cr}\right)\,,
\ee
are complimentary projectors on 
$L^2(W\oplus W)$, i.e.
\be
B^2=B \qquad K^2=K, \qquad BK=KB=0\,.
\ee
Hence, a section that satisfies the boundary conditions can be
pa\-ra\-met\-ri\-zed by $\chi(\varphi)=u(\varphi)\oplus v(\varphi) \in
L^2(W\oplus W)$,  $u(\varphi)\in L^2_{\perp}$, $v(\varphi)\in L^2_{||}$,
so that
\be
\psi(\varphi)=K\chi(\varphi)=\left(\matrix{u(\varphi)\cr
-\Lambda u(\varphi)+v(\varphi)\cr}\right)\,.
\ee

It is not difficult to see that the boundary operator $B$ (\ref{5ms})
incorporates all standard types of boundary conditions. Indeed, by choosing
$\Pi=\II$ and $\Lambda=0$ one gets the Dirichlet boundary conditions, by
choosing $\Pi=0$,  $\Lambda=\II$ one gets the Neumann boundary conditions.
More generally, the choice $\Pi,\Lambda\in C^\infty(\End(W))$, so that 
$\Lambda\Pi=\Pi\Lambda=0$,  corresponds to the mixed boundary conditions.

\begin{rem}
The boundary $\partial M$ could  be, in general, a disconnected manifold
consisting of a finite  number of disjoint connected parts, $\partial 
M=\cup_{i=1}^n \Sigma_i$, with each $\Sigma_i$ being compact connected
manifold without boundary, $\partial\Sigma_i=  \emptyset$  and
$\Sigma_i\cap\Sigma_j= \emptyset$ if $i\ne j$. Thus one can  impose {\it
different} boundary conditions on different connected  parts of the boundary
$\Sigma_i$. This means that the full  boundary operator decomposes
$B=B_1\oplus \cdots \oplus B_n$,  with $B_i$ being different boundary
operators acting on different  bundles. 
\end{rem}

We always assume the manifold $M$ itself and the coefficients of the operator
$F$ to be smooth in the interior of $M$. If, in addition, the boundary 
$\partial M$ is smooth, and the boundary operator $B$ is a differential
operator  with smooth coefficients, then $(F,B)$ is called {\it smooth local 
boundary value problem}. 

In this paper we are interested in a different class of boundary conditions.
Namely, we do not assume the boundary operator to be smooth. Instead, we will
study the case when it has {\it discontinuous} coefficients.  Such problems
are often called mixed boundary conditions; to avoid misunderstanding we
will not use this terminology. We impose {\it different} boundary conditions
on {\it connected}  parts of the boundary, which makes the boundary value
problem {\it  discontinuous}. Roughly speaking, one has a decomposition of a
smooth boundary in some parts where {\it different types} of the boundary
conditions are imposed, i.e. Dirichlet or  Neumann. The boundary operator
is then {\it discontinuous} at  the intersection of these parts. The boundary
value problems of this type are called Zaremba problem in the literature
\cite{bruening85,bruening91} (see also 
\cite{sneddon66,fabrikant91,avramidi00,dowker00a,dowker00b}).

In this paper we consider the simplest case when there are just  two components.
We assume that the boundary of the manifold $\partial M$ is decomposed as the
{\it disjoint} union  
\be
\partial M=\Sigma_1\cup\Sigma_2\cup\Sigma_0\,,
\ee
where $\Sigma_1$ and $\Sigma_2$ are smooth compact submanifolds  of dimension
$(m-1)$ (codimension $1$ submanifolds), with the {\it same boundary}
$\Sigma_0=\partial\Sigma_1=\partial\Sigma_2$, that is a smooth compact
submanifold of dimension $(m-2)$ (codimension $2$ submanifold) without
boundary , i.e. $\partial \Sigma_0=\emptyset$. Let us stress here that when
viewed as sets both $\Sigma_1$ and $\Sigma_2$ are considered to be {\it
disjoint open sets}, i.e. $\Sigma_1\cap \Sigma_2=\emptyset$.

Let $\chi_i: \partial M\to {\mathbb R}$, $(i=0,1,2)$, be the characteristic
functions of the sets $\Sigma_i$, $\chi_i(\hat x)= 1$ if $\hat x\in \Sigma_i$
and $\chi_i(\hat x)=0$ if $\hat x \not\in\Sigma_i$. Obviously,
$\chi_1(\hat x)+\chi_2(\hat x)+\chi_0(\hat x)=1$ for any $\hat x\in\partial
M$. Let $\pi_i: L^2(W)\to L^2(W)$, $(i=0,1,2)$, be the trivial projections of
sections, $\psi$, of a vector bundle $W$ to $\Sigma_i$ defined by 
$(\pi_i\psi)(\hat x)=\chi_i(\hat x)\psi(\hat x)$, i.e. $(\pi_i\psi)(\hat x)=
\psi(\hat x)$ if $\hat x\in \Sigma_i$ and $(\pi_i\psi)(\hat x)=0$ if $\hat x
\not\in \Sigma_i$. In other words $\pi_1$ maps smooth sections of the bundle
$W$ to their restriction to $\Sigma_1$, extending them by zero on $\Sigma_2$,
and similarly for $\pi_2$. Obviously, $\pi_1+\pi_2+\pi_0=\Id$,
$\pi_i^2=\pi_i$, $(i=0,1,2)$, and $\pi_i\pi_j=0$ for $i\ne j$. Let $\Lambda\in
C^\infty(\End(W))$ be a smooth Hermitian endomorphism of the vector bundle $W$.
Then the boundary operator of our problem can be written in the
form   
\be 
B=\left(\matrix{\pi_1&0\cr
\pi_2\Lambda\pi_2 & \pi_2\cr}\right)\,,
\label{18ms}
\ee
The projectors $\pi_1$ and $\pi_2$ as well as the boundary operator $B$ are
clearly non-smooth ({\it discontinuous}) on $\Sigma_0$. In other words, we
have Dirichlet boundary conditions on $\Sigma_1$ and Neumann (Robin) boundary
conditions on $\Sigma_2$:  
\bea
&&\varphi\Big|_{\Sigma_1}=0,
\label{19ms}\\
&&(\nabla_N+\Lambda)\varphi\Big|_{\Sigma_2}=0 
\label{20ms}
\eea
Note that the boundary condition are set only on {\it open} subsets
$\Sigma_1$ and $\Sigma_2$; the boundary conditions do not say anything about
the boundary data on $\Sigma_0$. 
We will see later that to specify the solution uniquely we also need a
further condition which specifies the type of the singularity on $\Sigma_0$.

We will call the boundary value problem $(F,B)$ for a Laplace operator $F$ 
(\ref{1ms}) with the boundary operator $B$ of the form 
(\ref{18ms}) {\it Zaremba boundary value problem}.

\subsection{Symmetry}

Let us define the antisymmetric bilinear form
\be
I(\varphi_1,\varphi_2)\equiv 
(F\varphi_1,\varphi_2)_{L^2(V)}-(\varphi_1,F\varphi_2)_{L^2(V)}\,,
\label{27}
\ee
for any two smooth sections $\varphi_1,\varphi_2\in C^\infty(V)$ of the
vector bundle $V$. By integrating by parts on $M$ one can easily see that
this bilinear form depends only on the boundary data 
\be
I(\varphi_1,\varphi_2)=(\psi(\varphi_1),J\psi(\varphi_2))_{L^2(W\oplus W)}\,,
\ee
where
\be
J=\left(
\begin{array}{cc}
0    & \II\\
-\II & 0\\
\end{array}\right)\,.
\ee
Therefore, it vanishes on sections of the bundle $V$ with compact support
{\it disjoint} from the boundary $\partial M$ when the boundary data
vanish $\psi(\varphi_1)=\psi(\varphi_2)=0$. This is a simple consequence of
the fact that the operator $F$ is {\it formally self-adjoint}. A formally
self-adjoint operator  is {\it essentially self-adjoint} if its closure is
self-adjoint. This means that  the operator is such that: i) it is
{\it symmetric} on smooth sections satisfying the boundary conditions,
and ii) there exists a unique self-adjoint extension of it. To prove the
latter property one has to study the deficiency indices; however, this will
not be the subject of primary interest in the present paper. We check only the
first property, i.e. that the operator $F$ is symmetric.

By integrating by parts on $\partial M$, it is not difficult to check that the
form $I(\varphi_1,\varphi_2)$ does vanish for any $\varphi_1,\varphi_2 \in
C^\infty(V)$ satisfying the boundary conditions with the boundary operator
(\ref{5ms}) provided the operator $\Lambda$ is symmetric. Therefore, we 
immediately obtain that the  
Zaremba boundary value problem is symmetric.


\subsection{Ellipticity}

Let $F_0$ be a Laplace type operator  with {\it constant} coefficients
obtained from the operator $F$ by {\it freezing} the coefficients at a fixed
point $x_0$ in the leading derivative part, i.e. for the Laplace type operator
$F_0=-g^{\mu\nu}(x_0)\partial_\mu\partial_\nu$. If the point $x_0$ is in the
interior of the manifold $M$, we assume the operator $F_0$ to act on the
sections of the vector bundle $V$ over $\RR^m$. If the point $x_0$ is on the
boundary, we assume the operator $F_0$ to act on sections of $V$ over
$\RR^{m-1}\times\RR_+$. If the point $x_0$ is on the boundary, we also define
the operator $B_0$ by omitting the non-diagonal part of $B$ and 
freezing the coefficients of the diagonal part of
the boundary operator $B$,  i.e. $B_0=\chi_{1}(\hat x_0)\oplus
\chi_{2}(\hat x_0)$. Note that the coefficients of the operator $B_0$
are  {\it constant} on $\Sigma_1$ and $\Sigma_2$ but are {\it discontinuous}
on $\Sigma_0$.

The boundary value problem $(F,B)$ is said to be {\it elliptic with respect to
$\CC\setminus \RR_+$} if for any complex number $\lambda\in \CC\setminus \RR_+$
not lying on the positive real axis the following two conditions are
satisfied: i) for any interior point $x_0$ the equation 
\be
(F_0-\lambda)\varphi=0
\ee 
has a unique non-trivial solution  vanishing at infinity, and ii) for
any boundary point $\hat x_0$ the above equation has a unique
non-trivial solution vanishing at infinity and satisfying the boundary
conditions 
\be
B_0\psi(\varphi)=0\,.
\ee
Here non-trivial solution means that it is not identically zero, $\varphi\ne
0$, for $\lambda\ne 0$ and it is not constant, $\varphi\ne{\const}$, in
case $\lambda=0$.

The question of ellipticity of Zaremba boundary value problem is a subtle one.
We will show below that for the freezed problem to have a unique solution 
one has to impose an additional condition along the codimension two
submanifold $\Sigma_0$, which specifies the (singular) behavior of  he solution
near $\Sigma_0$. Without this condition (which is often imposed implicitly
by choosing the most regular solution) the freezed problem has infinitely
many acceptable (square integrable) solutions, so that Zaremba problem fails
to be elliptic.


\section{Heat Kernel}

For $t>0$ the heat semi-group  operator $U(t)=\exp(-tF):\ L^{2}(V,M)\to
L^2(V,M)$ is well defined. The kernel of this operator, called the heat
kernel,  is defined by the equation
\be
(\partial_t+F)U(t|x,x')=0
\ee
with the initial condition
\be
U(0|x,x')=\delta(x,x'),
\ee
where $\delta(x,x')$ is the covariant Dirac distribution, the boundary
condition 
\be
B\psi[U(t|x,x')]=0.
\ee
and the self-adjointness condition
\be
U(t|x,x')=U^*(t|x',x).
\ee
Hereafter all differential operators as well  as the boundary data map act on
the {\it first} argument of the heat kernel,  unless otherwise stated.

Let $\lambda$ be a complex number with a sufficiently large negative real
part, ${\rm Re}\, \lambda << 0$. The resolvent can then be defined by the
Laplace transform 
\be
G(\lambda)=\int\limits_0^\infty dt\, e^{t\lambda}U(t),
\label{3.8nnn} 
\ee
and by analytical continuation elsewhere. The heat kernel can be expressed,
in turn, in terms of the resolvent by the inverse Laplace transform
\be
U(t)={1\over 2\pi i}\int\limits_{w-i\infty}^{w+i\infty}d\lambda\,
e^{-t\lambda}G(\lambda)\,, 
\ee 
where $w$ is a sufficiently large negative real number, $w<<0$.  As it has
been done here, we will sometimes omit the space arguments if it does not
cause any confusion.

It is well known \cite{gilkey95} that  the heat kernel $U(t|x,x')$ is a
smooth function near diagonal of $M\times M$, i.e. for $x$ close to $x'$, and
has a well defined diagonal value  
\be
U^{\rm diag}(t|x)=U(t|x,x),
\ee
 and the functional trace
\be
\Tr_{L^2}\exp(-tF)=\int\limits_M \tr_V U^{\rm diag}(t)\,,
\ee
where $\tr_V$ is the fiber trace and the integration is defined with the help
of the usual Riemannian volume element $d\vol_g$.

It is also well known that in the smooth category  the trace of the heat
kernel has an asymptotic expansion   as $t\to 0^+$ of the form \cite{gilkey95}
\be
\Tr_{L^2}\exp(-tF)\sim \sum\limits_{k=0}^\infty
t^{(k-m)/2}A_{k}\,.
\ee
Here $A_{k}$ are the famous so-called ({\it global}) heat-kernel coefficients
(sometimes called also Minakshisundaram-Plejel coefficients).
They have the following general form \cite{gilkey95}: 
\be
A_{2k}=\int\limits_M a^{(0)}_{2k}
+\int\limits_{\partial M}a^{(1)}_{2k},
\ee
\be
A_{2k+1}=\int\limits_{\partial M}a^{(1)}_{2k+1},
\ee
where $a^{(0)}_{k}$ and $a^{(1)}_{k}$ are the ({\it local}) {\it interior}
and {\it boundary}  heat-kernel coefficients. The local interior coefficients
$a^{(0)}_k$ are also called {\it HMDS (Hadamard-Minackshisundaram-De
Witt-Seeley) coefficients} in the literature.  Hereafter the integration over
the boundary is defined with the help of the usual Riemannian volume element
$d\vol_{\hat g}$ on $\partial M$ with the help of the induced metric $\hat g$.

The interior coefficients $a^{(0)}_{k}$ do {\it not} depend on the boundary
conditions $B$. The even order coefficients $a^{(0)}_{2k}$ are  calculated
for Laplace-type operators up to $a^{(0)}_8$ \cite{avramidi91b,vandeven98}. The
boundary coefficients $a^{(1)}_{k}$ do  depend on {\it both} the operator $F$
and the boundary operator $B$.  They are far more complicated because in
addition to the geometry of the manifold $M$ they depend essentially on the
geometry of the  boundary $\partial M$. For Laplace-type operators they are
known  for the usual boundary conditions (Dirichlet,  Neumann, or mixed
version of them) up to $a^{(1)}_{5}$  \cite{branson90a,branson99,kirsten98}.
For oblique boundary conditions including tangential derivatives some
coefficients were recently computed in
\cite{mcavity91,avramidi99a,avramidi99b,avramidi98b,dowker97,dowker98}.

However, the boundary value problem considered in the present paper  with
Zaremba type boundary operator B (\ref{18ms}) is {\it essentially
singular}.
Even if the manifold $M$, its boundary $\partial M$  and the
operator $F$ are all in  smooth category, the coefficients of the  boundary
operator $B$ are discontinuous on $\Sigma_0$, which makes it a {\it singular
problem}. For such problems the asymptotic expansion of the trace of the heat
kernel has additional {\it non-trivial logarithmic terms},
\cite{gil01,bruening91}
\be
\Tr_{L^2}\exp(-tF_B)
\sim 
\sum\limits_{k=0}^{\infty} t^{(k-m)/2}B_{k}
+\log t\sum\limits_{k=0}^\infty t^{k/2}\,H_{k}.
\label{32ms}
\ee
Whereas there are some results concerning the coefficients $B_{k}$, almost
nothing is known about the coefficients $H_{k}$. Since the Zaremba problem
is local, or better to say `pseudo-local', all these coefficients have  the
form  
\be  B_{2k}=\int\limits_M b^{(0)}_{2k}
+\int\limits_{\Sigma_1}b^{(1),1}_{2k} +\int\limits_{\Sigma_2}b^{(1),2}_{2k}
+\int\limits_{\Sigma_0}b^{(2)}_{2k},
\label{38ms}
\ee
\be
B_{2k+1}=\int\limits_{\Sigma_1}b^{(1),1}_{2k+1}
+\int\limits_{\Sigma_2}b^{(1),2}_{2k+1}
+\int\limits_{\Sigma_0}b^{(2)}_{2k+1},
\label{39ms}
\ee
\be
H_{k}=\int\limits_{\Sigma_0}h_k\,.
\ee
Here the new feature is the appearance of the integrals over $\Sigma_0$, which
complicates the problem even more, since the coefficients now depend on the
geometry of the imbedding of the codimension $2$ submanifold $\Sigma_0$ in
$M$ that could be pretty complicated, even if smooth. The asymptotic expansion
of the trace of the heat kernel has been studied recently in \cite{seeley01}.
It has been shown there that the logarithmic terms do not appear, i.e.
$H_k=0$ for any $k$,  in the 
Zaremba type problem considered in the present paper, which confirmed the 
conjecture of \cite{avramidi00}.

\section{Parametrix of the Heat Equation}

Let us stress here that we are not going to provide a rigorous construction
of the parametrix with all  the estimates, which, for a singular 
boundary-value problem,  is a task that would require a separate paper.  For
such a treatment the reader is referred to the
papers
\cite{seeley69a,seeley69b,gilkey83b,gilkey83a,
booss-bavnbek93,grubb96}
for the smooth case and to \cite{fedosov64,cheeger79,cheeger83,callias83,
bruening85,bruening91,simanca87,mooers96,lesch97,karol98,gil01}  for
the singular case.

Here we shall adhere instead to a pragmatic approach and  will describe the
construction of the parametrix that can be used to calculate {\it explicitly}
the heat kernel coefficients $B_k$ as well as $H_k$.

\subsection{Geometrical Framework}

First of all, we need to describe properly the geometry of the problem. Let
us fix two small positive numbers $\varepsilon_1,\varepsilon_2>0$. We split
the whole manifold in a disjoint union of {\it four} different parts: 
\be 
M=M^{\rm int}\cup M^{\rm bnd}=
M^{\rm int}\cup M^{\rm bnd}_1\cup M^{\rm bnd}_2 \cup M^{\rm bnd}_0\,.
\ee
Here $M^{\rm bnd}_0$ is defined as the set of points in the narrow strip 
$M^{\rm bnd}$ of the manifold $M$ near the boundary $\partial M$ of the width
$\varepsilon_1$ that are at the same time in a narrow strip of the width
$\varepsilon_2$ near $\Sigma_0$ 
\be
M^{\rm bnd}_0=\{x\in M\ |\ {\rm dist}(x,\partial M)<\varepsilon_1, 
\ {\rm dist}(x,\Sigma_0)<\varepsilon_2\}\,.
\ee
Further, $M^{\rm bnd}_1$ is the part of the thin strip $M^{\rm bnd}$ of the
manifold  $M$ (of the width $\varepsilon_1$) near the boundary $\partial M$ 
that is near $\Sigma_1$ but on the finite distance from $\Sigma_0$, i.e.
\be
M^{\rm bnd}_1=\{x\in M\ |\ {\rm dist}(x,\Sigma_1)<\varepsilon_1, \ 
{\rm dist}(x,\Sigma_0)>\varepsilon_2\}\,.
\ee
Similarly,
\be
M^{\rm bnd}_2=\{x\in M\ |\ {\rm dist}(x,\Sigma_2)<\varepsilon_1, \ 
{\rm dist}(x,\Sigma_0)>\varepsilon_2\}\,.
\ee
Finally, 
$M^{\rm int}$ is the interior of the manifold $M$ without a thin strip at the
boundary $\partial M$, i.e.
\be
M^{\rm int}=M\setminus \left(M^{\rm bnd}_1\cup M^{\rm bnd}_2 \cup M^{\rm
bnd}_0\right)
=\{x\in M\ |\ {\rm dist}(x,\partial M)>\varepsilon_1\}\,.
\ee

We will construct the parametrix on $M$ by using {\it different approximations}
in different domains. Strictly speaking, to glue them together in a smooth way 
one should use `smooth characteristic functions' of different domains
(partition of unity) and carry out all necessary estimates. What one has to
control is the order of the remainder terms in the limit $t\to 0$ and
their dependence on $\varepsilon_1$ and $\varepsilon_2$.  Since our task here
is not to prove the form of the asymptotic expansion (\ref{32ms}), which is
known, but rather to compute explicitly the coefficients of the asymptotic
expansion, we will not worry about such subtle details. We will compute the
asymptotic expansion as $t\to 0$ in each domain and then take the limit
$\varepsilon_1,\varepsilon_2\to 0$. For a rigorous treatment see
\cite{bruening85,bruening91,gil01} and the references therein.   

We will use different local coordinates in different domains.
In $M^{\rm int}$ we do not fix the local coordinates; 
our treatment will be manifestly covariant. 

In $M^{\rm bnd}_1$ we choose the local coordinates as follows. Let
$\{\hat e_i\}$, $(i=2,\dots m)$, be the local frame for the
tangent bundle  $T\Sigma_1$ and $\hat x=(\hat x^i)=(\hat x^2,\dots,\hat
x^{m})$,  $(i=2,\dots,m)$, be the local coordinates on $\Sigma_1$.  Let $r={\rm
dist}(x,\Sigma_1)$ be the normal distance  to $\Sigma_1$, ($r=0$ being the
defining equation of $\Sigma_1$), and $\hat N=\partial_r\big|_{\Sigma_1}$ be
the inward pointing unit normal to $\Sigma_1$.  Then by using the geodesic flow
we get the local frame $\{N(r,\hat x),e_i(r,\hat x)\}$ 
for the tangent bundle $TM$ and the local
coordinates $x=(r,\hat x)$  on $M_1^{\rm bnd}$. The geometry of $\Sigma_1$ is
described by the extrinsic curvature $K$ (second fundamental form)
\be
\hat \nabla_i e_j=K_{ij} N\,,\qquad
\hat\nabla_i N=-K^j{}_i e_j\,.
\ee

The coordinate $r$ ranges from $0$ to $\varepsilon_1$, $0\le
r\le\varepsilon_1$. The local coordinates in $M_2^{\rm bnd}$ are chosen
similarly.

Finally, in $M^{\rm bnd}_0$ we choose the local coordinates as follows.
Let $\{\hat e_a(\hat x)\}$, $(a=3,\dots,m)$, be a local frame for
the  tangent bundle $T\Sigma_0$ and let $\hat x=(\hat x^a)=(\hat x^3,\dots,\hat
x^{m})$ be the local coordinates on $\Sigma_0$. To avoid misunderstanding we
should stress here that now we use the same notation $\hat x$ to denote
coordinates on $\Sigma_0$ (not on the whole $\partial M$).  Let
$\dist_{\partial M}(x,\Sigma_0)$ be the distance from a point $x$ on
$\partial M$ to $\Sigma_0$ {\it along the boundary} $\partial M$. Then define
$y=+\dist_{\partial M}(x,\Sigma_0)>0$ if $x\in \Sigma_1$ and
$y=-\dist_{\partial M}(x,\Sigma_0)<0$ if $x\in \Sigma_2$. In other
words, $y=0$ on $\Sigma_0$, ($r=y=0$ being the defining equations of
$\Sigma_0$), $y>0$ on $\Sigma_1$ and $y<0$ on $\Sigma_2$. Let $\hat n(\hat
x)=\partial_y\big|_{\Sigma_0}$ be the unit normal to $\Sigma_0$
pointing inside $\Sigma_1$. Then by using the {\it tangential geodesic flow}
along the boundary (that is normal to $\Sigma_0$) we first get the local
orthonormal frame $\{n(y,\hat x),e_a(y,\hat x)\}$ for the tangent bundle
$T\partial M$. Further, let the unit normal vector field to the boundary $\hat
N(y,\hat x)$ be defined as above. Then by using the {\it normal geodesic flow}
to the boundary we get the local frame $\{N(r,y,\hat x),n(r,y,\hat
x),e_a(r,y,\hat x)\}$ for the tangent bundle $TM$ and local coordinates
$(r,y,\hat x)$ on $M_0^{\rm bnd}$. The geometry of $\Sigma_0$ (codimension $2$
manifold) is described by {\it two extrinsic curvatures} $K$ and $L$ and an
additional vector $T$:   \be
\hat\nabla_a e_b=K_{ab}n+L_{ab}N\,. 
\ee
\be
\hat \nabla_a n=-K^b{}_a e_b+T_a N\,,
\qquad
\hat \nabla_a N=-L^b{}_a e_b-T_a n\,.
\ee


The ranges of the coordinates $r$ and $y$ are:
$0\le r\le\varepsilon_1$ and $-\varepsilon_2\le y\le\varepsilon_2$.
Finally, we introduce the polar coordinates 
\be
r=\rho\cos\theta, \qquad
y=\rho\sin\theta\,.
\ee
The angle $\theta$ ranges from $-\pi/2$ to $\pi/2$ with 
$\theta=-\pi/2$ on $\Sigma_1$ and $\theta=\pi/2$ on $\Sigma_2$. To cover the
whole $M^{\rm bnd}_0$, $\rho$ should range from $0$ to some $\varepsilon_3$
(depending on $\varepsilon_1$ and $\varepsilon_2$),
$0\le\rho\le\varepsilon_3$.

\subsection{Interior Parametrix}

This is the easiest case. The construction of the parametrix goes along the
same lines as for manifolds without boundary (see, e.g.
\cite{dewitt84,gilkey95,avramidi91b,avramidi96,avramidi99}. The basic case
(when the coefficients of the operator $F$ are frozen at a point   $x_0$) is,
in fact, {\it zero-dimensional}, i.e.  algebraic. By using the normal
coordinates at $x_0$ and Fourier transform one easily obtains the {\it leading
 heat kernel}  \be
U^{\rm int}_0(t|x,x')=(4\pi t)^{-m/2}\exp\left(-{|x-x'|^2\over 4t}\right)\,.
\ee

We try to find the fundamental solution of the heat equation near diagonal for
small $t$, i.e. $x\to x'$ and $t\to 0^{+}$, that, instead of the boundary
conditions satisfies asymptotic condition at infinity. This means that
effectively one introduces a small expansion parameter  $\varepsilon$
reflecting the fact that the points $x$ and $x'$  are close to each other and
the parameter $t$ is small. This can be done by fixing a point $x_0=x'$ in
$M^{\rm int}$, choosing the normal  coordinates at this point (with
$g_{\mu\nu}(x')=\delta_{\mu\nu}$), scaling 
\be
x\to x'+\varepsilon(x-x'), \qquad y\to x'+\varepsilon(y-x'), 
\qquad t\to \varepsilon^2t,
\ee
and expanding in a power series in $\varepsilon$.
We will 
label the scaled objects by $\varepsilon$, e.g.
$U^{\varepsilon}$. The scaling parameter  $\varepsilon$ will be considered as
a small parameter in the theory and we  will use it to expand everything in
power (asymptotic) series in $\varepsilon$. At the very end of calculations we
set $\varepsilon=1$.  The non-scaled objects, i.e. those with $\varepsilon=1$,
will not have the label $\varepsilon$.  Another way of doing this is by saying
that we will  expand all quantities in the homogeneous functions of
$(x-x')$, $(y-y')$ and $\sqrt t$. This construction is
standard and we do not repeat it here.

One can also use instead a manifestly covariant method
\cite{dewitt84,avramidi96,avramidi91b,avramidi98,avramidi99,vandeven98}, which
gives a convenient formula for the asymptotics as $t\to 0^{+}$ 
\be 
U^{\rm int}(t)\sim
\exp\left(-{\sigma\over 2t}\right) \Delta^{1/2}
\sum\limits_{k=0}^\infty {t^{(k-m)/2}}a_k,
\label{3.19xxx}
\ee
where $\sigma=\sigma(x,x')=(1/2)[\dist(x,x')]^2$ is one half of the square
of the geodesic distance between $x$ and $x'$, $\Delta=\Delta(x,x')=g^{-1/2}$
$(x)g^{-1/2}(x')$ $\det[-\partial^x_\mu\partial^{x'}_{\nu}\sigma(x,x')]$ is
the corresponding Van Vleck-Morette determinant, $g=\det g_{\mu\nu}$,  and
$a_k=a_k(x,x')$ are the off-diagonal heat-kernel coefficients (note that odd
order coefficients vanish identically, i.e.  $a_{2k+1}=0$). These coefficients
satisfy certain  differential recursion relations which can be solved in form
of a covariant Taylor series near diagonal \cite{avramidi91b}.

The asymptotic expansion of the heat kernel on the diagonal reads
\be
U^{\rm int}_{\rm diag}(t)
\sim \sum\limits_{k=0}^\infty t^{(k-m)/2}a^{\rm diag}_{k}\,,
\label{48-nm}
\ee
where $a_k^{\rm diag}(x)=a_k(x,x)$. This asymptotic
expansion can be integrated over the interior of the manifold $M^{\rm int}$.
Since both the local interior coefficients $a_k$ and the volume element
$d\vol_g$ are regular at the boundary,  these integrals have well defined
limits as $\varepsilon_1\to 0$ 
\be
\lim_{\varepsilon_1\to 0^+}
\int\limits_{M^{\rm int}}\tr_V a^{\rm diag}_k=\int\limits_M\tr_V a^{\rm diag}_k\,.
\ee
Thus we obtain the {\it local interior} contribution to the global heat kernel
coefficients $B_k$:
\be
b^{(0)}_{2k}=\tr_V a^{\rm diag}_{2k}\,.
\ee

As we already noted above all odd order coefficients vanish,
$b_{2k+1}^{(0)}=0$. The explicit formulas for even order
coefficients $b_{2k}^{(0)}$ are known up  to $b_{8}^{(0)}$
\cite{avramidi91b,vandeven98}.  The first two coefficients have the
well known form  
\be 
b_0^{(0)}=(4\pi)^{-m/2}\dim V\,,
\ee
\be
b_{2}^{(0)}=(4\pi)^{-m/2}\tr_V\left(Q-{1\over 6}R\right)\,,
\ee
where $R$ is the scalar curvature.

\subsection{Dirichlet Parametrix}

In this section we will follow closely the ideas of the paper \cite{avramidi93}.
For an {\it elliptic} boundary-value problem  the diagonal of the parametrix
$U^{\rm diag}_{\rm bnd}(t)$ in $M^{\rm bnd}_1$ has  {\it exponentially small}
terms, i.e. of order $\sim\exp(-r^2/t)$,  (recall that $r$ is the normal
geodesic distance to the boundary) as $t\to 0^{+}$ and $r>0$. These terms do
not contribute to the asymptotic expansion of the heat-kernel diagonal outside
the boundary as $t \to 0^{+}$. However, they behave like {\it distributions
near the boundary}, and, therefore, the integrals over $M_1^{\rm bnd}$, more
precisely, the integrals
$\lim_{\varepsilon_1\to 0}\int\limits_{\Sigma_1}\int\limits_0^{\varepsilon_1}dr(\dots)$, do
contribute to the asymptotic expansion with coefficients being the integrals
over $\Sigma_1$. It is {\it this phenomenon} that leads to the boundary terms
in the heat kernel coefficients. Thus, such terms determine the local boundary
contributions  $b^{(1)}_{k}$ to the global heat-kernel coefficients $B_{k}$.
The same applies to the Neumann parametrix and $\Sigma_2$.

The Dirichlet parametrix $U^{\rm bnd, (1)}(t|x,x')$ in $M^{\rm bnd}_1$ is
constructed as follows.   Now we want to find the fundamental solution of the
heat equation near diagonal, i.e. for $x\to x'$ and for small $t\to 0$ in the
region $M^{\rm bnd}_{1}$ {\it close to the boundary}, i.e. for  small $r$ and
$r'$, that satisfies Dirichlet boundary conditions on $\Sigma_1$ and
asymptotic condition at infinity. We fix a point on the boundary,
$x_0\in\Sigma_1$, and choose normal coordinates on $\Sigma_1$  at this point
(with $g_{ij}(0,\hat x_0)=\delta_{ij}$). 

The basic case here (when the coefficients of the operator $F$ are  frozen at
the point $x_0$ is {\it one-dimensional}.  The zeroth-order term $U^{\rm
bnd,(1)}_{0}$ is defined by the heat equation
\be
(\partial_t+F_0)U^{\rm bnd,(1)}_{0}=0,
\ee
where
\be
F_0=-\partial_r^2-\hat\partial^2\,,
\label{59ms}
\ee
the initial condition 
\be
U^{\rm bnd,(1)}_0(0|r,\hat x;r',\hat x')=\delta(r-r')\delta(\hat x,\hat x'),
\ee
the boundary conditions, 
\be
U^{\rm bnd,(1)}_{0}\Big|_{\Sigma_1}=0,
\ee
and the asymptotic condition
\be
\lim_{r\to\infty} U^{\rm bnd,(1)}_{0}(t|r,\hat x;r',\hat x')=
\lim_{r'\to\infty} U^{\rm bnd,(1)}_{0}(t|r,\hat x;r',\hat x')=0\,.
\ee
Note that the restriction to the boundary $(\dots)\Big|_{\Sigma_1}$ applies
only to the first argument, i.e. $r\to 0$. The operator $F_0$ is a partial
differential operator with {\it constant coefficients}. By using the Fourier
transform in the boundary coordinates  $(\hat x-\hat x')$ it reduces to an
{\it ordinary} differential operator of second order. Clearly, the $\Sigma_0$
part factorizes and the solution to the remaining one-dimensional problem can
be easily obtained by using the Laplace transform, for example. The leading
order Dirichlet parametrix thus has the form 
\be
U^{\rm bnd,(1)}_{0}(t|r,\hat x;r',\hat x')=
K(t|r,\hat x;r',\hat x')-
K(t|r,\hat x;-r',\hat x')
\ee
where
\be
K(t|r,\hat x;r',\hat x')=
(4\pi t)^{-m/2}\exp\left(-{|\hat x-\hat x'|^2+(r-r')^2\over 4t}\right)\,,
\label{64ms}
\ee
Note that in addition to the usual symmetry of the heat
kernel, the Dirichlet parametrix possesses the following 
`{\it mirror symmetry}'
\bea
\lefteqn{U^{\rm bnd,(1)}_{0}(t|r,\hat x;r',\hat x')}
\nonumber\\
&&=-U^{\rm bnd,(1)}_{0}(t|-r,\hat x;r',\hat x')
=-U^{\rm bnd,(1)}_{0}(t|r,\hat x;-r',\hat x')\,,
\eea
i.e. it is an {\it odd} function of the coordinates $r$ and $r'$ separately.

To construct the whole parametrix, we again scale the coordinates. But now we
include the coordinates $r$ and $r'$ in the scaling
\be
\hat x\to \hat x_0+\varepsilon(\hat x-\hat x_0),\qquad 
\hat x'\to \hat x_0+\varepsilon(\hat x'-\hat x_0)
\ee
\be
r\to \varepsilon r,\qquad 
r'\to \varepsilon r', \qquad
t\to \varepsilon^2 t\,.
\ee
The corresponding differential operators are scaled by
\be
\hat\partial\to{1\over\varepsilon}\hat\partial, \qquad
\partial_r\to{1\over\varepsilon}\partial_r,\qquad
\partial_t\to{1\over\varepsilon^2}\partial_t\,.
\ee
Then, we expand the scaled operator $F_\varepsilon $ in the power 
series in $\varepsilon$, i.e.
\be
F\to F_\varepsilon\sim \sum\limits_{n=0}^\infty\varepsilon^{n-2} F_n,
\ee
where $F_n$ are second-order differential operators with homogeneous symbols.
Since the Dirichlet boundary operator does not contain any derivatives and has
constant coefficients on $\Sigma_1$ it does not scale at all.

The subsequent strategy is rather simple.  We expand the scaled heat kernel
in $\varepsilon$ 
\be
U^{\rm bnd,(1)}_{\varepsilon}\sim\sum_{n=0}^\infty
\varepsilon^{2-m+n}U^{\rm bnd,(1)}_{n},
\ee
and substitute into the scaled version of the heat equation and the Dirichlet
boundary condition on $\Sigma_1$.  Then, by equating  the like powers in
$\varepsilon$ one gets an infinite set of recursive differential equations 
\be
(\partial_t+F_0)U^{\rm bnd,(1)}_{k}
=-\sum\limits_{n=1}^{k}F_nU^{\rm bnd,(1)}_{k-n}, \qquad
k=1,2,\dots,
\ee
with the boundary conditions
\be
U^{\rm bnd,(1)}_{k}(t|0,\hat x;r',\hat x')=
U^{\rm bnd,(1)}_{k}(t|r,\hat x;0,\hat x')=0,
\ee
and the asymptotic conditions
\be
\lim_{r\to\infty} U^{\rm bnd,(1)}_{k}(t|r,\hat x;r',\hat x')=
\lim_{r'\to\infty} U^{\rm bnd,(1)}_{k}(t|r,\hat x;r',\hat x')=0\,.
\label{asymp-inf}
\ee

In other words, we decompose the parametrix into the homogeneous parts with
respect to $(\hat x-\hat x_0)$, $(\hat x'-\hat x_0)$, $r$, $r'$ and 
$\sqrt t$, i.e.
\bea
\lefteqn{U^{\rm bnd,(1)}_{k}(t|r,\hat x;r',\hat x')}
\nonumber\\
&&=t^{(k-m)/2} U^{\rm bnd,(1)}_{k}\left(1|t^{-1/2}r,\hat x'
+t^{-1/2}(\hat x-\hat x');
t^{-1/2}r',\hat x'\right)\,,
\eea
in particular, on the diagonal we have
\bea
U^{\rm bnd,(1)}_{k}(t|r,\hat x;r,\hat x)
=t^{(k-m)/2} U^{\rm bnd,(1)}_{k}\left(1|t^{-1/2}r,\hat x;
t^{-1/2}r,\hat x\right)\,,
\eea
and, therefore,
\be
U^{\rm bnd,(1)}_{\rm diag}(t)
\sim
\sum_{k=0}^\infty t^{(k-m)/2}
U^{\rm bnd,(1)}_{k}(1|t^{-1/2}r,\hat x;t^{-1/2}r,\hat x)\,.
\ee
To compute the contribution to the asymptotic expansion of the trace of the 
heat kernel, we will need to compute the integral of 
$U^{\rm bnd,(1)}_{\rm diag}(t)$ over $M_1^{\rm bnd}$.
One should stress that the volume element should also be scaled 
\be
d\vol(r,\hat x)\to d\vol(\varepsilon r,\hat x) 
=d\vol(0,\hat x)\cdot\sum_{k=0}^\infty \varepsilon^k {r^k\over k!}g_k(\hat
x)\,  
\ee
where
\be
g_k(\hat x)=
{\partial^k\over\partial r^k}\left[{d\vol(r,\hat x)\over d\vol(0,\hat
x)}\right]\Bigg|_{r=0}\,.
\ee
Combining the above equations and changing the variable 
$r=\sqrt{t}\xi$ we obtain
\bea
&&\int\limits_{M_1^{\rm bnd}} 
U^{\rm bnd, (1)}_{\rm diag}(t)
=\int\limits_{\Sigma_1}
\int\limits_0^{\varepsilon_1} dr\,
{d\vol(r,\hat x)\over d\vol(0,\hat x)}
U^{\rm bnd, (1)}_{\rm diag}(t|r,\hat x;r,\hat x)
\nonumber\\[5pt]
&&
\sim
\sum_{k=0}^\infty t^{(k-m)/2}
\int\limits_{\Sigma_1}
\sum_{n=0}^{k-1}{1\over n!} g_n(\hat x)
\int\limits_0^{\varepsilon_1/\sqrt{t}} d\xi\, 
\xi^n
U^{\rm bnd,(1)}_{k-n-1}(1|\xi,\hat x;\xi,\hat x)\,,
\eea

We note that even if the coefficients $U_k^{\rm bnd,(1)}$ satisfy the asymptotic
regularity condition at $r\to\infty$ (\ref{asymp-inf}) off-diagonal, 
the diagonal values of them do not fall off at infinity. 
They have the following general form
\be
U_k^{\rm bnd, (1)}(1|\xi,\hat x;\xi,\hat x)
=P_k(\xi,\hat x)
+Y_k^{(1)}(\xi,\hat x)\,,
\ee
where $P_k(\xi,\hat x)$ are {\it polynomials} in $\xi$ and 
$Y_k^{(1)}(\xi,\hat x)$ are {\it exponentially small}, more precisely
$\sim \xi^{\alpha}\exp(-\xi^2)$ with some $\alpha$, as $\xi\to\infty$
(which corresponds to $t\to 0$).

Obviously, the integrals over the polynomial part over $M^{\rm bnd}_1$
vanish after taking the asymptotic expansion as $t\to 0$ and the limit 
${\varepsilon_1,\varepsilon_2\to 0}$. The coefficients $P_k$ constitute
simply
the `interior part' of the parametrix and are not essential in computing the
boundary contribution.
The coefficients $Y_k^{(1)}$, in contrary, 
behave like distributions near $\Sigma_1$. They 
give the $\Sigma_1$ contributions to the boundary heat kernel coefficients
$b_k^{(1)}$. 
In the limit
$t\to 0$ the integral $\int_0^{\varepsilon_1/\sqrt t}d\xi(\dots)$ 
becomes  
$\int\limits_0^{\infty}d\xi\,(\dots)$ plus an exponentially small remainder
term. Then in the limit $\varepsilon_1\to 0$ we obtain integrals over
$\Sigma_1$ up to an exponentially small function that we are not interested
in. 

As the result we get the coefficients $b_k^{(1),1}$ in the form 
\be 
b_k^{(1),1}
=\sum_{n=0}^{k-1}{1\over n!}g_n\int\limits_0^\infty  d\xi\,\xi^n
\tr_V Y^{(1)}_{k-n-1}(\xi,\hat x)\,.
\ee
These are the standard boundary heat kernel coefficients for Dirichlet
boundary conditions. They are listed for example in
\cite{branson90a,branson99} up to $k=4$. The first two have the form
\bea
b_0^{(1),1}&=&0\,,
\nonumber\\
b_1^{(1),1}&=&-(4\pi)^{-(m-1)/2}\,\dim V\,{1\over 4}\,,
\nonumber\\
b_2^{(1),1}&=&(4\pi)^{-m/2}\,\dim V\, {1\over 3}\, K\,,
\eea
where $K$ is the trace of the extrinsic curvature (second fundamental form) of
the boundary.

\subsection{Neumann Parametrix}

The construction of the Neumann parametrix in $M^{\rm bnd}_2$ is essentially
the same except that now the boundary operator, in fact the endomorphism
$\Lambda$, is not constant and should be also scaled, so that the scaled
boundary conditions are \cite{avramidi99a,avramidi99b}  
\be 
\left({1\over\varepsilon}\partial_r
+\Lambda_\varepsilon\right)\varphi\Big|_{\Sigma_2}=0\,,
\ee
where 
\be
\Lambda_\varepsilon\sim\sum\limits_{k=0}^\infty \varepsilon^{k} \Lambda_{k}\,.
\ee
The zeroth-order operator $F_0$ is given by the same formula (\ref{59ms})
and the zero order boundary operator is just the standard Neumann one.
The basic zero-order problem can again be easily solved by
\be
U^{\rm bnd,(2)}_{0}(t|r,\hat x;r',\hat x')=
K(t|r,\hat x;r',\hat x')+
K(t|r,\hat x;-r',\hat x')
\ee
with the same kernel $K$ (\ref{64ms}).
Note that the Neumann parametrix has another
{\it mirror symmetry}
\bea
\lefteqn{U^{\rm bnd,(2)}_{0}(t|r,\hat x;r',\hat x')}
\nonumber\\
&&=U^{\rm bnd,(2)}_{0}(t|-r,\hat x;r',\hat x')
=U^{\rm bnd,(2)}_{0}(t|r,\hat x;-r',\hat x')\,,
\eea
i.e. it is an {\it even} function of the coordinates $r$ and $r'$ separately.

The construction of the parametrix goes along the same lines as in Dirichlet
case. We have the recursive differential equations
\be
(\partial_t+F_0)U^{\rm bnd, (2)}_{k}=-\sum\limits_{n=1}^{k}F_n
U^{\rm bnd,(2)}_{k-n}, \qquad
k=1,2,\dots,
\ee
with the boundary conditions
\be
\partial_r U^{\rm bnd,(2)}_{k}\Big|_{\Sigma_2}
=-\sum_{n=1}^{k-1} \Lambda_n U^{\rm bnd,(2)}_{k-n-1}\Big|_{\Sigma_2},
\ee
and the asymptotic conditions
\be
\lim_{r\to\infty} U^{\rm bnd,(2)}_{k}(t|r,\hat x;r',\hat x')=
\lim_{r'\to\infty} U^{\rm bnd,(2)}_{k}(t|r,\hat x;r',\hat x')=0\,.
\ee
As we already noted above the restriction to the boundary applies only to the
first argument $r$. One can repeat here everything said at the end of the
previous subsection about Dirichlet parametrix.  We have again homogeneity
property 
\bea
\lefteqn{U^{\rm bnd,(2)}_{k}(t|r,\hat x;r',\hat x')}
\nonumber\\
&&=t^{(k-m)/2} U^{\rm bnd,(2)}_{k}(1|t^{-1/2}r,\hat x'+t^{-1/2}(\hat x-\hat x');
t^{-1/2}r',\hat x')
\eea
and the following expansion for the diagonal
\be
U^{\rm bnd,(2)}_{\rm diag}(t)
\sim
\sum_{k=0}^\infty t^{(k-m)/2}
U^{\rm bnd,(2)}_{k}(1|t^{-1/2}r,\hat x;t^{-1/2}r,\hat x)\,.
\ee
By separating the polynomial and exponentially small parts,
\be
U_k^{\rm bnd, (2)}(1|\xi,\hat x;\xi,\hat x)
=P_k(\xi,\hat x)
+Y_k^{(2)}(\xi,\hat x)\,,
\ee
and repeating the arguments at the end of the previous subsection 
we obtain
the $\Sigma_2$ contributions to the boundary heat kernel coefficients
$b_k^{(1),2}$
\be
b_k^{(1),2}
=\sum_{n=0}^{k-1}{1\over n!}g_n\int\limits_0^\infty  d\xi\,\xi^n
\tr_V Y^{(2)}_{k-n-1}(\xi,\hat x)\,.
\ee
These are the standard boundary heat kernel coefficients for Neumann boundary
conditions.
They are listed for example in
\cite{branson90a,branson99} up to $k=4$. The first two have the form
\bea
b_0^{(1),2}&=&0\,,
\nonumber\\
b_1^{(1),2}&=&(4\pi)^{-(m-1)/2}\,\dim V\,{1\over 4}\,,
\nonumber\\
b_2^{(1),2}&=&(4\pi)^{-m/2}\,\dim V\, {1\over 3}\, K\,.
\eea

\subsection{Mixed Parametrix}

This is the most complicated (and the most interesting) case, since here the
basic problem with frozen coefficients on $\Sigma_0$ is {\it two-dimensional}.
More precisely, in $M^{\rm bnd}_0$ the basic problem is on the half-plane.  
Since the origin is a singular point, we will work in polar coordinates
introduced above. 

\subsubsection{Basic Problem (Zeroth Order)}

First of all, we need to solve the basic problem for operators with frozen
coefficients at a point $\hat x_0$ on $\Sigma_0$. We choose normal coordinates
on $\Sigma_0$ at this point (with $g_{ab}(0,\theta,\hat x_0)=\delta_{ab}$)
and the polar coordinates $(\rho,\theta)$ in the normal bundle described 
above.
Then the zero order operator $F_0$ has the form  
\be 
F_0=-\partial_\rho^2 - {1\over \rho}\partial_\rho  -{1\over \rho^2}
\partial_\theta^2-\hat\partial^2  \,
\ee
where $\hat\partial^2=\hat g^{ab}\hat\partial_a\hat\partial_b$.
The zero order inward pointing normal $N$ to the boundary 
in polar coordinates has the form
\bea
N_0\Big|_{\Sigma_1}
&=&\partial_r\Big|_{y>0}
=-{1\over \rho }\partial_\theta\Big|_{\rho>0,\; \theta={\pi\over 2}}\,,
\\[10pt]
N_0\Big|_{\Sigma_2}
&=&\partial_r\Big|_{y<0}
= {1\over \rho }\partial_\theta\Big|_{\rho>0,\; \theta=-{\pi\over 2}}\,,
\\[10pt]
N_0\Big|_{\Sigma_0}
&=&\partial_r\Big|_{y=0}
=\partial_\rho\Big|_{\rho=0,\;\theta=0}\,.
\eea
Now the boundary operator is discontinuous, and there  is a {\it singularity}
at the origin $\rho=0$.

Again the part due to $\Sigma_0$ factorizes
\bea
\lefteqn{U^{\rm bnd,(0)}_0(t|\rho,\theta,\hat x;\rho',\theta',\hat x')}
\nonumber\\
&&=(4\pi t)^{-(m-2)/2}\exp\left(-{|\hat x-\hat x'|^2\over 4t}\right) 
\Psi(t|\rho,\theta;\rho',\theta')\,,
\eea
where
$\Psi(t|\rho,\theta;\rho',\theta')$ is a
two-dimensional heat kernel.
It is determined by the heat equa\-tion   
\be
\left(\partial_t -\partial_\rho^2-{1\over\rho}\partial_\rho
-{1\over\rho^2}\partial_{\theta}^2\right)
\Psi(t|\rho,\theta;\rho',\theta') = 0 \,,
\ee
the initial condition
\be
\Psi(0^+|\rho,\theta;\rho',\theta') = {1\over \sqrt{\rho\rho'} }\,
\delta(\rho-\rho')\delta(\theta-\theta')\,, 
\ee
the boundary conditions
\be
\Psi(t|\rho,\theta;\rho',\theta')\Big|_{\theta={\pi\over 2}}=0\,,
\ee
\be
\partial_\theta
\Psi(t|\rho,\theta;\rho',\theta')\Big|_{\theta=-{\pi\over 2}}=0\,, 
\ee
the symmetry condition
\be
\Psi(t|\rho,\theta;\rho',\theta')
=\Psi(t|\rho',\theta';\rho,\theta)\,,
\ee
as well as some boundary conditions at $\rho \to 0^+$ and $\rho\to\infty$.

We require certain regularity conditions at infinity, 
\be
\int\limits_0^\infty d\rho\,\sqrt{\rho\rho'}\,
\left|\Psi(t|\rho,\theta;\rho',\theta')\right|<\infty\,,
\ee
in particular,
\be
\lim_{\rho\to\infty}\sqrt{\rho\rho'}\,\Psi(t|\rho,\theta;\rho',\theta')=
\lim_{\rho\to\infty}\partial_\rho\left[\sqrt{\rho\rho'}\,
\Psi(t|\rho,\theta;\rho',\theta')\right]=0\,. 
\ee

As far as the boundary condition on $\Sigma_0$, i.e. at
$\rho=0$, is concerned, we 
will see that it cannot be a generic condition, rather, like
for the usual Frobenius theory of differential equations near singular points,
one has a couple of possibilities for the type of singularity that need to be
specified.

Since we are looking for the solution of the heat equation whose
diagonal is integrable near boundary, we require
\be
\int\limits_{0}^{\varepsilon_3} d\rho\,\rho\,
|\Psi(t|\rho,\theta;\rho,\theta)| < \infty\,.
\ee
This effectively imposes a restriction on the type of the singularity at
$\rho\to 0$, i.e. the singularity of the heat kernel at $\rho\to 0$ must be
weaker than $(\rho\rho')^{-1}$. More precisely, we assume that
\be
\sqrt{\rho\rho'}\,|\Psi(t|\rho,\theta;\rho',\theta')| < \infty\,.
\ee

We will see that this is still 
not enough to fix a unique solution and another
boundary condition at $\rho\to 0$ is needed. Since the point $\rho=0$
is singular, this boundary condition cannot be imposed arbitrarily.
Also it does not follow from the boundary conditions on $\Sigma_1$ 
and $\Sigma_2$. We impose it in one of the following forms
\be
\left.
\left[\sqrt{\rho\rho'}\,
\Psi(t|\rho,\theta;\rho',\theta')\right]\right|_{\rho=0^+}
=0\,,
\label{sing-dir}
\ee
or
\be
\left.(\partial_\rho-s)
\left[\sqrt{\rho\rho'}\,
\Psi(t|\rho,\theta;\rho',\theta')\right]\right|_{\rho=0^+}
=0\,,
\label{sing-neumn}
\ee
where $s$ is a real parameter.
We will see that the heat kernel asymptotics do depend on  this 
boundary condition as well.  We call the boundary condition (\ref{sing-dir})
``regular'' boundary condition. It corresponds formally to the limit
$s\to \infty$.

To construct the heat kernel we study first the operator
\be
L=-\partial_\theta^2 
\label{106mn}
\ee 
on the interval $[-{\pi\over 2},{\pi\over 2 }]$ with the boundary conditions
\be
\varphi(\theta)\Big|_{\theta={\pi\over 2}}=0, \qquad
\partial_\theta\varphi(\theta)\Big|_{\theta=-{\pi\over 2}}=0\,.
\ee 
It is not difficult to find the spectral resolution of this  operator. Its
orthonormal eigenfunctions and eigenvalues are 
\be
\varphi_n(\theta)=\sqrt{{2\over\pi}}
\cos\left[\left(n+{1\over 2
}\right)\left(\theta+{\pi\over 2}\right)\right]\,,
\ee
\be
\lambda_n=\left(n+{1\over 2}\right)^2\,,
\label{ln}
\ee 
where $n=0,1,2,\dots$.

By separating the variables
\be
\Psi(t|\rho,\theta;\rho',\theta')=\sum_{n=0}^{\infty}
\varphi_n(\theta)\varphi_n(\theta')u_n(t|\rho;\rho') 
\ee 
we obtain the equation
\be
\left[\partial_t
-\partial^2_\rho-{1\over \rho}\partial_\rho
+{1\over \rho^2}\left(n+{1\over 2}\right)^2
\right]u_n(t|\rho;\rho') = 0\,,
\label{he-n}
\ee
with the initial condition
\be
u_n(0^+|\rho;\rho') = {1\over \sqrt{\rho\rho'} } \delta(\rho-\rho'),
\label{init-n}
\ee
the symmetry condition
\be
u_n(t|\rho;\rho')=u_n(t|\rho';\rho)\,
\ee
and the asymptotic conditions at infinity
\be
\int\limits_0^\infty d\rho\,\sqrt{\rho\rho'}\,
\left|u_n(t|\rho,\theta;\rho',\theta')\right|<\infty\,,
\ee
\be
\lim_{\rho\to\infty}\sqrt{\rho\rho'}\,u_n(t|\rho;\rho')=
\lim_{\rho\to\infty}\partial_\rho\left[
\sqrt{\rho\rho'}\,u_n(t|\rho;\rho')\right]=0\,.
\label{asymp-n}
\ee
The boundary conditions at $\rho=0$ are
\be
\sqrt{\rho\rho'}|u_n(t|\rho,\rho')|<\infty\,,
\label{int-n}
\ee
and, more precisely, one of the following
\be
\left.
\left[\sqrt{\rho\rho'}\,
u_n(t|\rho,\theta;\rho',\theta')\right]\right|_{\rho=0^+}
=0\,,
\label{sing-dir-n}
\ee
or
\be
\left.(\partial_\rho-s)
\left[\sqrt{\rho\rho'}\,
u_n(t|\rho,\theta;\rho',\theta')\right]\right|_{\rho=0^+}
=0\,.
\label{sing-neumn-n}
\ee

Let us consider the operator
\be
D_n=-\partial^2_\rho-{1\over \rho}\partial_\rho
+{1\over \rho^2}\left(n+{1\over 2}\right)^2\,.
\ee
It has the ``eigenfunctions'' $J_\nu(\mu \rho)$:
\be
D_n J_{\nu}(\mu \rho)
=\mu^2 J_{\nu}(\mu \rho)\,,
\label{eigen-n}
\ee
where $\mu$ is a positive real parameter, 
$J_\nu(z)$ are Bessel functions of the first kind of order $\nu$,
and $\nu$ can take one of two values, either $\nu=\left(n+{1\over 2}\right)$
or $\nu=-\left(n+{1\over 2}\right)$. 
However, the behavior at $\rho\to 0$ of the Bessel function
$J_{-\left(n+{1\over 2}\right)}(\mu\rho)$ 
for $n\ge 1$ is too singular,
$\sim \rho^{-\left(n+{1\over 2}\right)}$, which
violates the integrability condition (\ref{int-n}).
This means that for any $n\ge 1$ we have to choose
$
\nu=\left(n+{1\over 2}\right)\,.
$
Note that these are not ``true'' eigenfunctions,
since they are non-normalizable.
{}Rather they satisfy the following ``orthogonality'' condition
\be
\int\limits_0^\infty d\mu \,\mu\,J_{n+{1\over 2}}(\mu\rho)
J_{n+{1\over 2}}(\mu\rho')=
{1\over\sqrt{\rho\rho'}}\delta(\rho-\rho')\,.
\ee

In the case $n=0$ both choices are possible, i.e. 
$\nu=+1/2$ or $\nu=-1/2$,
which makes the analysis of the problem more complicated.
Therefore, we will treat the cases $n\ge 1$ and $n=0$ separately.

\bigskip
{\sc Case I.}
We consider first the case $n\ge 1$.
We will solve this problem by employing the Hankel transform 
which is well defined in the class of functions 
satisfying the conditions imposed above. We define
\be
v_n(t|\mu,\rho')=\int\limits_0^\infty d\rho\,\rho 
J_{n+{1\over 2}}(\mu\rho)u_n(t|\rho,\rho')\,,
\ee
then
\be
u_n(t|\rho,\rho')=\int\limits_0^\infty d\mu\,\mu\, 
J_{n+{1\over 2}}(\mu\rho)v_n(t|\mu,\rho')\,,
\ee

Next, by integrating by parts and using the eq. (\ref{eigen-n}), 
we compute the Hankel transform
\bea
\lefteqn{\int\limits_0^\infty d\rho\,\rho\, J_{n+{1\over 2}}(\mu\rho)D_n 
u_n(t|\rho,\rho')
=\mu^2 \int\limits_0^\infty d\rho\,\rho\, J_{n+{1\over 2}}(\mu\rho)
u_n(t|\rho,\rho')}
\nonumber\\
&&
+\left.\rho\left\{
\left[\partial_\rho J_{n+{1\over 2}}(\mu\rho)\right]u_n(t|\rho,\rho')
-J_{n+{1\over 2}}(\mu\rho)\partial_\rho u_n(t|\rho,\rho')
\right\}\right|_0^\infty\,.
\nonumber\\
\eea
Finally, by taking into account the boundary conditions (\ref{asymp-n}) and
(\ref{int-n}), and the asymptotic form of the Bessel functions,
we obtain
\be
\int\limits_0^\infty d\rho\,\rho\, J_{n+{1\over 2}}(\mu\rho)D_n 
u_n(t|\rho,\rho')
=\mu^2 v_n(t|\mu,\rho')\,.
\ee
Thus, the Hankel transform of the heat equation (\ref{he-n}) is
\be
(\partial_t+\mu^2) v_n(t|\mu,\rho')=0\,.
\ee
{}From (\ref{init-n}) we also obtain the initial condition
\be
v_n(0^+|\mu,\rho')=J_{n+{1\over 2}}(\mu\rho')\,.
\ee
It immediately follows that
\be
v_n(t|\mu,\rho')=e^{-t\mu^2}J_{n+{1\over 2}}(\mu\rho')\,,
\ee
and, therefore,
\be
u_n(t|\rho,\rho') = \int\limits_0^\infty d\mu\,\mu\,
e^{-t\mu^2}J_{n+1/2}(\mu\rho)J_{n+1/2}(\mu\rho')\,.
\label{un-sol0}
\ee
This integral can be computed by using the properties of the Bessel functions.
We obtain finally
\be
u_n(t|\rho,\rho') = {1\over 2t}
\exp\left(-{\rho^2+\rho'^2\over 4t}\right)
I_{n+1/2}\left({\rho\rho'\over 2t}\right)\,,
\label{un-sol}
\ee
where $I_{n+1/2}(z)$ is the modified Bessel function of first kind.
Note that this solution satisfies both boundary conditions (\ref{sing-dir-n})
and (\ref{sing-neumn-n}). 

\bigskip
{\sc Case II.} Now let us consider the case $n=0$.  As we have seen
the condition of integrability (\ref{int-n}) does not fix the solution
uniquely, since there are two linearly independent solutions that 
satisfy that condition, which corresponds to the choices $\nu=-1/2$ and
$\nu=+1/2$. The Hankel transform in this case reduces to the standard
cosine and sine Fourier transforms. However, we will not use them, but will
solve the heat equation directly. 

Let us single out the allowed singular factor
\be
u_0(t|\rho,\rho')={1\over\sqrt{\rho\rho'}}w(t|\rho,\rho')\,.
\ee
Then, the heat equation (\ref{he-n}), the initial condition (\ref{init-n}),
and the boundary conditions (\ref{sing-dir-n}) and (\ref{sing-neumn-n}) take
the form
\be
(\partial_t-\partial_{\rho}^2)w(t|\rho,\rho')=0\,,
\ee
\be
w(0^+|\rho,\rho')=\delta(\rho-\rho')
\ee
\be
w(t|\rho,\rho')\Big|_{\rho=0}=0,
\label{wdir}
\ee
or
\be
(\partial_\rho-s)w(t|\rho,\rho')\Big|_{\rho=0}=0\,.
\ee
There is also the usual regularity condition at infinity $\rho\to\infty$.

As we see, $w$ is just the standard one-dimensional heat kernel on the
half-axis. 
By using the Laplace transform we easily obtain the solution of this problem
\bea
w(t|\rho,\rho')
&=&{1\over 2\pi i}\int\limits_{c-i\infty}^{c+i\infty}
d\lambda\,e^{-t\lambda}{1\over 2\sqrt{-\lambda}}
\Bigg\{
\exp\left[{-\sqrt{-\lambda}|\rho-\rho'|}\right]
\nonumber\\[5pt]
&&
+{\sqrt{-\lambda}-s\over \sqrt{-\lambda}+s}
\exp\left[{-\sqrt{-\lambda}(\rho+\rho')}\right]
\Bigg\}\,,
\eea
where $c$ is a sufficiently large {\it negative} real constant, i.e.
$\sqrt{-c}>-s$, and 
$\sqrt{-\lambda}$ is defined in the complex plane of $\lambda$ with a cut along
the real positive half-axis, so that ${\rm Re}\,\sqrt{-\lambda}>0$.
Notice that the boundary conditions (\ref{wdir}) correspond to the limit
$s\to+\infty$. The limit $s\to-\infty$ is not well defined since the constant
$c$ depends on $s$ and would have to go to $-\infty$ as well. 

Next, let us change the variable $\lambda$ according to
\be
\lambda = \mu^2, \qquad
\mu=i\sqrt{-\lambda}\,,
\ee
where ${\rm Im}\,\mu>0$. In the upper
half-plane, ${\rm Im}\,\mu>0$, this change of variables is single-valued
and well defined. Under this change the
complex $\lambda$-plane is mapped onto the upper half $\mu$-plane, and
the cut in the complex $\lambda$-plane
along the positive real axis from $0$ to $\infty$ is mapped onto the
whole real axis in the $\mu$-plane.

The contour of integration in the complex $\mu$-plane
is a hyperbola going from $(e^{i3\pi/4})\infty$ through the point 
$\sqrt{-c}$ to
$(e^{i\pi/4})\infty$. It   
can be deformed to a contour $C$ that 
is above all poles of the integrand.
It comes from 
$-\infty$ along the real axis, encircles posible poles
on the imaginary axis in the clockwise direction, 
and goes to $+\infty$ along the real axis.

After such a transformation we obtain
\bea
w(t|\rho,\rho')&=&
\int\limits_{C} 
{d\mu\over 2\pi}\, \Bigg\{\exp\left[{-t\mu^2+i\mu|\rho-\rho'|}\right]
\nonumber\\
&&
+{\mu-is\over\mu+is}\exp\left[{-t\mu^2+i\mu(\rho+\rho')}\right]\Bigg\}.
\label{c0}
\eea
This function is an analytic function of $s$ since the contour $C$ is above 
the pole at $-is$. Therefore, we can compute it, for example, for $s>0$,
and then make an analytical continuation on the whole complex $s$-plane.
So, let $s>0$. Then the pole $-is$ is in the lower half-plane.
Therefore, the contour $C$ can be deformed to just the real axis, i.e.
$-\infty<\mu<\infty$. Next, we use
the following trick
\be
{\mu-is\over\mu+is}=1-2is{1\over \mu+is}
=1-2s\int\limits_0^\infty dp\, e^{ip(\mu+is)}\,.
\ee
This integral converges since $s>0$. Substituting this equation in 
(\ref{c0}) and evaluating the Gaussian integral over $\mu$, we obtain
\bea
w(t|\rho,\rho')&=&(4\pi t)^{-1/2}\Bigg\{
\exp\left[-{(\rho-\rho')^2\over 4t}\right]
+\exp\left[-{(\rho+\rho')^2\over 4t}\right]
\nonumber\\
&&
-2s\int\limits_0^\infty dp\,\exp\left[-{(\rho+\rho'+p)^2\over 4t}-ps\right]
\Bigg\}\,,
\eea
which can be expressed in terms of the complimentary error function
\bea
\lefteqn{
w(t|\rho,\rho')=(4\pi t)^{-1/2}\Bigg\{
\exp\left[-{(\rho-\rho')^2\over 4t}\right]
+\exp\left[-{(\rho+\rho')^2\over 4t}\right]}
\nonumber\\[5pt]
&&
-2\sqrt{\pi}s\sqrt{t}\exp\left[{ts^2+(\rho+\rho')s}\right]{\rm erfc}\,\left(
{\rho+\rho'\over 2\sqrt{t}}+s\sqrt{t}\right)
\Bigg\}\,.
\label{w0}
\eea
Here ${\rm erfc}\,(z)$ is defined by
\be
{\rm erfc}\,(z)={2\over\sqrt{\pi}}\int\limits_z^\infty du\,e^{-u^2}\,.
\ee
The case $s<0$ can be analyzied either directly or by the analytical 
continuation. The direct computation is different since now the pole
$-is$ is in the upper half-plane and one has to take into account
the residue at this pole. However, the integral along the real axis 
is also different, so that the sum is the same. In other words, the result for
$s<0$ has the same analytical form (\ref{w0}).

Finally, we obtain the heat kernel component $u_0$:
\bea
&&
u_0(t|\rho,\rho')=(4\pi t)^{-1/2}{1\over \sqrt{\rho\rho'}}\Bigg\{
\exp\left[-{(\rho-\rho')^2\over 4t}\right]
+\exp\left[-{(\rho+\rho')^2\over 4t}\right]
\nonumber\\[5pt]
&&\qquad
-2\sqrt{\pi}s\sqrt{t}\exp\left[{ts^2+(\rho+\rho')s}\right]{\rm erfc}\,\left(
{\rho+\rho'\over 2\sqrt{t}}+s\sqrt{t}\right)
\Bigg\}\,.
\label{u0}
\eea
In the particular case $s=0$ we get
\be
u_0(t|\rho,\rho')=(4\pi t)^{-1/2}{1\over \sqrt{\rho\rho'}}\left\{
\exp\left[-{(\rho-\rho')^2\over 4t}\right]
+\exp\left[-{(\rho+\rho')^2\over 4t}\right]
\right\}\,.
\label{u0neu}
\ee
The case $s\to+\infty$ corresponds to the regular boundary conditions
(\ref{sing-dir-n}). In this case the solution reads
\bea
u_0(t|\rho,\rho')
&=&
(4\pi t)^{-1/2}{1\over \sqrt{\rho\rho'}}\left\{
\exp\left[-{(\rho-\rho')^2\over 4t}\right]
-\exp\left[-{(\rho+\rho')^2\over 4t}\right]\right\}
\nonumber\\[10pt]
&=& {1\over 2t}
\exp\left(-{\rho^2+\rho'^2\over 4t}\right)
I_{1/2}\left({\rho\rho'\over 2t}\right)\,,
\label{u0dir}
\eea
and coincides with the solution (\ref{un-sol}) for $n=0$
obtained by the Hankel transform.

Combining our results and  using the explicit form of the
eigenfunctions $\varphi_n$, we obtain the heat kernel
\bea
\lefteqn{
\Psi(t|\rho,\theta;\rho',\theta')
=
(4\pi t)^{-1}\Phi(t|\rho,\rho')}
\nonumber\\[5pt]
&&\times
\left\{\cos\left({\theta-\theta'\over 2}\right)
+ \cos\left({\theta+\theta'+\pi\over 2} \right)
\right\}
\nonumber\\[5pt]
&&
+(4\pi t)^{-1} 
\exp\left(-{\rho^2+\rho'^2\over 4t}\right)
\nonumber\\
&&\times
\left\{\Omega\left({\rho\rho'\over 2t}, \theta-\theta'\right)+
\Omega\left({\rho\rho'\over 2t},\theta+\theta'+\pi\right)\right\}
\,,
\eea 
where
\bea
\lefteqn{\Phi(t|\rho,\theta;\rho',\theta')
={4\over\sqrt{\pi}}\left({t\over \rho\rho'}\right)^{1/2}
\Bigg\{\exp\left[-{(\rho+\rho')^2\over 4t}\right]}
\nonumber\\[5pt]
&&
-\sqrt{\pi}s\sqrt{t}\exp\left[{ts^2+(\rho+\rho')s}\right]{\rm erfc}\,\left(
{\rho+\rho'\over 2\sqrt{t}}+s\sqrt{t}\right)
\Bigg\}\,,
\eea
and
\be
\Omega(z,\gamma) = 2\sum_{n=0}^\infty
I_{n+1/2}(z) \cos \left[ \left(n+{1\over 2 } \right)\gamma\right] \,. 
\ee 
Notice that for the ``regular'' boundary conditions (\ref{sing-dir}), which 
correspond to the limit $s\to+\infty$, the function
$\Phi(t|\rho,\rho')$ vanishes.

This series can be evaluated 
by using the following integral representation of the Bessel
function 
\be
I_{n+1/2}(z)={1\over \sqrt{\pi}\,n!}\left({z\over
2}\right)^{n+1/2} \int\limits_{-1}^1 dp\, e^{-pz}(1-p^2)^{n}\,.
\ee
Substituting this integral in the series and summing over $n$ we obtain
\bea
\Omega(z,\gamma) &=& \sqrt{{z\over 2\pi}} \int\limits_{-1}^1 dp\,e^{-pz}
\Biggl\{\exp\left[{1\over 2}(1-p^2)ze^{i\gamma} 
+ {1\over 2 }i\gamma \right] 
\nonumber\\
&&
+ \exp\left[{1\over 2}(1-p^2)ze^{-i\gamma} -
{1\over 2 }i\gamma \right] \Biggr\}\,. 
\eea 
The remaining integral can be expressed in terms of the error function, so
that finally we get
\be
\Omega(z,\gamma)=e^{z\cos\gamma} 
\erf\left[\sqrt{2z}\,\cos\left({\gamma\over 2}\right)\right] 
\ee 
where the error function is defined by
\be
\erf(z)={2\over \sqrt\pi}\int\limits_0^z dp\, e^{-p^2} \,.
\ee

By adding the $\Sigma_0$ factor we obtain the final answer for the
parametrix
\be
U^{\rm bnd,(0)}_0(t|\rho,\theta;\rho',\theta')=
L(t|\rho,\theta,\hat x;\rho',\theta',\hat x')
+L(t|\rho,\theta,\hat x;\rho',-\theta'-\pi,\hat x')
\ee
where
\bea
\lefteqn{
L(t|\rho,\theta,\hat x;\rho',\theta',\hat x')}
\nonumber\\
&&=
(4\pi t)^{-m/2}\exp\left(-{|\hat x-\hat x'|^2\over 4t}\right)
\Phi(t|\rho,\rho')\cos\left({\theta-\theta'\over 2}\right)
\nonumber\\
&&+
(4\pi t)^{-m/2}
\exp\left\{-{1\over 4t}\left[{|\hat x-\hat x'|^2+\rho^2+\rho'^2-2\rho\rho'
\cos(\theta-\theta')}\right]\right\} 
\nonumber\\
&&\times
\erf\left(\sqrt{\rho\rho'\over t}
\cos\left({\theta-\theta'\over 2}\right)\right)\,.
\eea

An important corollary from this formula are the symmetries of the heat kernel.
First of all, we have the usual `self-adjointness' symmetry
\be
\theta \to \theta',\qquad \rho\to \rho'\,.
\ee
Second, we have the `periodicity' symmetries
\be
\theta\to \theta +4\pi n, \qquad \theta'\to\theta'+4\pi m,
\qquad n,m\in\ZZ\,.
\ee
Finally, there is additional `mirror' symmetry
\bea
&&\theta\to \theta, \qquad \theta'\to-\theta'-\pi \\
&&\theta\to -\theta-\pi, \qquad \theta'\to \theta'
\eea
Note the essential {\it difference of the symmetries} of the mixed parametrix
versus those of the Dirichlet and Neumann parametrices. The mixed parametrix
is a periodic function of the angles (expected), but not with the period
$2\pi$ but with the period $4\pi$ (not expected). That is why there are
two {\it different mirror images}, $(\rho,-\theta-\pi,\hat x)$ and
$(\rho,-\theta+\pi,\hat x)$,  of a point with the coordinates
$(\rho,\theta,\hat x)$. In other words the double reflection of a point
does not bring it back---{\it the double image is not identical with the original
point}. Denoting by $T$ the transformation $\theta\to -\theta-\pi$ we have
\be
T^4=\Id, \qquad {\rm but} \qquad T^2\ne \Id\,.
\ee
The operator $T$ has {\it four} eigenvalues $1$, $-1$, $i$ and $-i$. Whereas
the first two, $1$ and $-1$, are the standard ones, the latter two, $i$ and
$-i$, correspond to some {\it new} images. This might have some interesting
applications.

The diagonal of the mixed parametrix is easily found to
be
\bea
\lefteqn{
U^{\rm bnd,(0)}_{{\rm diag},0}(t)
=
(4\pi t)^{-m/2}
\Bigg\{1-{\rm erfc}\,\left({\rho\over \sqrt{t}}\right)
}
\nonumber\\[5pt]
&&
-\exp\left(-{\rho^2\cos^2\theta\over t}\right)
{\rm erf}\,\left({\rho\sin\theta\over\sqrt{t}}\right)
\nonumber\\[5pt]
&&
+
(1-\sin\theta) {4\over \sqrt{\pi}} {\sqrt{t}\over\rho}
\Bigg[\exp\left(-{\rho^2\over t}\right)
\nonumber\\[5pt]
&&
-\sqrt{\pi}s\sqrt{t}\exp\left({ts^2+2\rho\,s}\right)
{\rm erfc}\,\left({\rho\over \sqrt{t}}+s\sqrt{t}\right)
\Bigg]
\Biggr\}\,. 
\label{131ms}
\eea

Now we compute the integral of the diagonal of the parametrix over
$M^{\rm bnd}_0$
\be 
\int\limits_{M^{\rm bnd}_0} \tr_VU^{\rm bnd,(0)}_{{\rm diag},0}(t) 
=\int\limits_0^{\varepsilon_3} d\rho\, \rho 
\int\limits_{-\pi/2}^{\pi/2}d\theta
\tr_V U^{\rm bnd,(0)}_{{\rm diag},0}(t)
\ee
for some finite $\varepsilon_3>\sqrt{\varepsilon^2_1+\varepsilon_2^2}>0$.

First of all, 
obviously the integrals over $\theta$ of the odd functions in $\theta$
vanish identically. So, we only need to consider the even part.
Second, since in the limit $\varepsilon_3\to 0$ the volume of $M^{\rm bnd}_0$
vanishes, the regular part of the heat kernel diagonal does not contribute to
the trace either. It is only the singular part of the heat kernel diagonal, 
which behaves like a distribution near $\Sigma_0$, that contributes to the
integral in the limit $\varepsilon_3\to 0$.

The integral over $\rho$ can be computed exactly. It reads
\bea
\int\limits_{M^{\rm bnd}_0} \tr_VU^{\rm bnd,(0)}_{{\rm diag},0}(t)
&=&
\int\limits_{\Sigma_0}
(4\pi t)^{-m/2}\dim V
\Biggl\{
{\pi\varepsilon_3^2\over 2}
\nonumber\\
&&
+t\left[-{\pi\over 4}+2\pi\Theta\left(\sqrt{t}s\right)\right]
+X(t)
\Biggr\}\,,
\eea
where
\be
\Theta(z)=e^{z^2}\erfc(z)\,,
\ee
and
\bea
X(t)
&=&
2\sqrt{\pi t}\,\varepsilon_3
\exp\left(-{\varepsilon_3^2\over t}\right)
+\left({\pi t\over 4}-{\pi\varepsilon_3^2\over 2}\right)
\erfc\left({\varepsilon_3\over\sqrt{t}}\right)
\nonumber\\
&&
-2\pi t \exp\left(ts^2+2s\varepsilon_3\right)
\erfc\left({\varepsilon_3\over \sqrt t}+\sqrt{t}s\right)\,.
\eea
Notice that $\pi\varepsilon_3^2/2$ is nothing but the area of the semi-circle
of radius $\varepsilon_3$, so that $\vol(\Sigma_0) \pi\varepsilon_3^2/2=
\vol(M_0^{\rm bnd})$. In the limit $\varepsilon_3\to 0$ this term does not
contribute to the asymptotics.

By using the asymptotic behavior of the error function
as $z\to\infty$
\bea
\erfc(z)&\sim& {1\over \sqrt\pi z}\,e^{-z^2}\,.
\label{asymp-error}
\eea
we find that the function $X(t)$ is exponentially small, i.e. it is suppressed
by the factor $\sim\exp(-\varepsilon_3^2/t)$,
as $t\to 0$,
and, therefore, does not contribute to the asymptotic expansion of the
heat kernel in powers of $t$ (\ref{32ms}) either.

The behavior of the function $\Theta(\sqrt{t}s)$ depends on the parameter $s$. 
For a finite
$s$ in the limit $t\to 0$ we have
\be
\Theta\left(\sqrt{t}s\right)= 1+O(t^{1/2})\,.
\ee
It immediately follows that for a finite $s$, i.e. for the boundary conditions
(\ref{sing-neumn}) the singular heat kernel coefficient $b_2^{(2)}$
is equal to
\be
b_2^{(2)}=(4\pi)^{-(m-2)/2}\dim V\,{7\over 16}\,.
\ee 
Notice that it does not depend on $s$ explicitly.

Finally we analyze the regular boundary conditions (\ref{sing-dir}), which
corresponds formally to the limit $s\to +\infty$. By using 
(\ref{asymp-error}) we see that for a finite $t$ 
the function $\Theta(\sqrt{t}s)$ vanishes in the limit $s\to\infty$:
\be
\Theta\left(\sqrt{t}s\right)\Big|_{s\to\infty}=0\,.
\ee
Therefore, in this case the coefficient $b_2^{(2)}$ is
\be
b_2^{(2)}=-(4\pi)^{-(m-2)/2}\dim V\,{1\over 16}\,.
\ee

We see that the heat kernel coefficients $b_{k}^{(2)}$ depend on the
{\it type} of the additional boundary conditions, i.e. (\ref{sing-dir}) vs.
(\ref{sing-neumn}), at $\Sigma_0$.


\section{Conclusions}

We have studied Zaremba boundary value problem for second-order 
partial differential operators of Laplace type, when the manifold as well
as the boundary are smooth and the differential operator has smooth
coefficients but the boundary operator is discontinuous on the boundary,
it jumps from the Dirichlet to Neumann type boundary operator.
Since this problem is not smooth there could be additional logarithmic
terms in the asymptotic expansion of the trace of the heat kernel 
(\ref{32ms}). However, Seleey \cite{seeley01}
has shown recently that such terms do not appear and there is classical
asymptotic expansion in half-integer powers of $t$ only. This seems to 
contradict the conclusions of \cite{dowker00a}, where it has been shown
that such an expansion with {\it locally computable} coefficients does 
not exist. The term `locally computable' is confusing though. As we have seen
the calculation of the coefficients of the asymptotic expansion 
of the trace of the heat kernel involves the knowledge of 
some global information, i.e.
the spectrum (\ref{ln}) of the operator $L$ (\ref{106mn}) 
with mixed boundary conditions. So, one could say that these coefficients
are locally computable in the coordinates $\hat x$ and $\rho$ but are 
{\it global} in the coordinate $\theta$. Thus the standard 
asymptotic expansion in powers of $t$ (without logarithmic terms) 
still exists with coefficients
(\ref{38ms}), (\ref{39ms}) 
given by integrals over $M$, $\Sigma_1$, $\Sigma_2$ and $\Sigma_0$.
The interior coefficients, $b_k^{(0)}$, the co-dimension one coefficients, 
$b_k^{(1),1}$ and $b_k^{(1),2}$, 
are `locally computable', but the co-dimension two coefficients, $b_k^{(2)}$,
are `global' in $\theta$ (or pseudo-local) 
and require new methods of calculation (e.g. like the approach of this
paper). They are constructed from the local invariants on $\Sigma_0$. It
is the numerical coefficients that are global.

Let us formulate briefly our main results. First of all, we provide
the correct formulation of the Zaremba type boundary value problem. We find
that the boundary conditions on the open sets $\Sigma_1$ and $\Sigma_2$ are
not enough to fix the problem, 
and an additional boundary condition along the singular set
$\Sigma_0$ is needed. This additional boundary condition can be considered
formally 
as an `extension' of Dirichlet conditions from $\Sigma_1$ to $\Sigma_0$,
or an `extension' of Neumann conditions from $\Sigma_2$ to $\Sigma_0$.
However, strictly speaking the boundary conditions on $\Sigma_0$ 
does not follow from the boundary conditions on $\Sigma\setminus\Sigma_0$ 
and can be chosen rather arbitrarily. One needs some additional `physical'
criteria to fix this boundary condition.
Second, we describe the geometry of the problem, which involves now some
nontrivial geometrical quantities 
(normal bundle and extrinsic curvatures) 
that characterize properly the imbedding of 
a co-dimension two submanifold $\Sigma_0$ in $M$. The higher order coefficients
$b_k^{(2)}$ are invariants constructed from those geometric quantities.
Next, we describe the construction of the parametrix of the heat equation
in the interior of the manifold $M$, in a thin shell close to $\Sigma_1$ and 
$\Sigma_2$, and finally, in a thin strip close to $\Sigma_0$.
We used the standard scaling device; the difference is just in what
coordinates are involved in scaling. Finally, we have explicitly found the 
off-diagonal parametrix in $M_0^{\rm bnd}$, the thin strip close to $\Sigma_0$,
in the leading approximation, and used it to compute the first nontrivial
`global' coefficient, $b_2^{(2)}$, of the heat kernel asymptotic expansion. 
We considered
two types of the additional boundary condition along $\Sigma_0$, one being
the `extension' Dirichlet boundary conditions (that we called regular boundary
condition), and another being the `extension' of the Neumann (or rather Robin)
boundary conditions. We have shown that the result, i.e. the coefficient 
$b_2^{(2)}$, does depend on the type of the boundary condition, i.e.
Dirichlet vs Neumann,
but does not depend on the parameter $s$ of the Robin boundary condition
(it will however contribute to the higher order coefficients).

\section*{Acknowledgements}
I would like to thank Jochen Br\"uning, Stuart Dowker, Giampiero Esposito, 
Stephen Fulling, Peter Gilkey, Gerd Grubb, and Werner M\"uller 
for stimulating and fruitful discussions. I am also very grateful to 
Robert Seeley for clarifying discussions of the boundary conditions  
and sharing the preliminary results. 
The support by the NSF Block Travel Grant DMS-9988119,
by the MSRI, and by the Istituto
Italiano per gli Studi Filosofici and the Azienda Autonoma Soggiorno e
Turismo, Napoli, is gratefully acknowledged.



\end{document}